\documentclass[prd,aps,floats,tabularx,preprintnumbers,twocolumn]{revtex4}
\usepackage{graphicx, epsfig}

\def\be{\begin{equation}}
\def\ee{\end{equation}}
\def\la{\label}
\def\bea{\begin{eqnarray}}
\def\eea{\end{eqnarray}}

\def\ci{\cite}
\def\la{\label}

\def\le{\left}
\def\ri{\right}
\def\fr{\frac}

\def\Omde{\Omega_{DE}}

\def\rp{\rho_\phi}
\def\rpo{\rho_{\phi o}}
\def\wp{w_\phi}

\def\Ompo{\Omega_{\phi o}}

\def\rim{\rho_{IM}}
\def\rimo{\rho_{IM o}}
\def\Oimo{\Omega_{IM o}}
\def\lmim{\lambda_{IM }}
\def\lmp{\lambda_{\phi }}

\begin{document}

\title{The impact of neutrino masses on the determination
of dark energy properties}

\author{Axel De La Macorra$^\sharp$, Alessandro Melchiorri$^{\flat}$,
Paolo Serra$^{\flat}$, Rachel Bean$^*$}

\affiliation{$^\sharp$ Instituto de F\'{\i}sica, UNAM, Apdo. Postal 20-364,
01000 M\'exico D.F., M\'exico\\
$^\flat$ Physics Department, University of Rome ``La Sapienza'' and
INFN - Sezione di Roma,
Ple Aldo Moro 2, 00185, Rome, Italy\\
$*$ Dept. of Astronomy, Cornell University, Ithaca, NY 14853\\}

\begin{abstract}
Recently, the Heidelberg-Moscow double
beta decay experiment has claimed a detection for a neutrino mass with high
significance. Here we consider the impact of this measurement on the
determination of the dark energy equation of state.
By combining the Heidelberg-Moscow result with the WMAP 3-years
data and other cosmological datasets we constrain the equation of
state to $-1.67< w <-1.05$ at $95 \%$ c.l., ruling out
a cosmological constant at more than $95 \%$ c.l..
Interestingly enough, coupled neutrino-dark energy models may
be consistent with such equation of state.
While future data are certainly needed for a confirmation of the
controversial Heildelberg-Moscow claim, our result shows that
future laboratory searches for neutrino masses may play a crucial
role in the determination of the dark energy properties.

\end{abstract}


\maketitle

The recent results of precision cosmology and the measurements of
Cosmic Microwave Background (CMB) anisotropies are in excellent
agreement with the standard model of structure formation (see e.g.
~\cite{wmap3cosm}). The price-tag of this success story is a very
puzzling consequence: the evolution of the universe is dominated by
a mysterious form of energy, coined dark energy "DE" (an unclustered
negative pressure component of the mass-energy density) with a
present-day energy density fraction $\Omega_{DE} \simeq 2/3$ and
equation of state $w \sim -1$ \cite{wmap3cosm}, \cite {seljakanze}.
This discovery may turn out to be one of the most important
contribution to physics in our generation.

Up to now, most data analysis have found a dark energy scenario
consistent with a true cosmological constant $\Lambda$, with $w=
-1$. However, the nature of dark matter and dark energy remains an
enigma, and we are entering an era when their origins might be
better understood not only through precision indirect
(observational) evidence from cosmological measurements but also,
crucially, through {\it direct} ground based measurements of
particles within and beyond the Standard model. In this {\it letter}
we discuss such a combination of data looking at what can be learned
about the possible interactions between neutrinos and dark energy
from combining cosmological observations with the neutrino mass
measurements from the Heidelberg-Moscow experiment
(\cite{Kl04},\cite{Kl06}, HM hereafter).

Let us recall that in the past years mass differences between neutrino mass eigenstates
($m_1$,$m_2$,$m_3$) have been measured in oscillation
experiments \cite{pastor}.
Observations of atmospheric neutrinos suggest a squared
mass difference of $\Delta m^2 \sim 3 \times 10^{-3} eV^2$ and
solar neutrino observations, together with results from the KamLAND
reactor neutrino experiment, point towards $\Delta m^2 \sim 5 \times
10^{-5} eV^2$. While only weak constraints on the absolute mass scale
($\Sigma m_{\nu} = m_1+m_2+m_3$) have been obtained from single $\beta$-decay experiments,
double beta decay searches from the HM experiment
have reported a signal for a neutrino mass
at $>4\sigma$ level \cite{Kl04}, recently
promoted to $>6\sigma$ level by a pulse-shape analysis \cite{Kl06}.
As we will see in the next section, this claim translates in a
total neutrino mass of $\Sigma m_{\nu} > 1.2 eV$ at $95 \%$ c.l..
While this claim is still considered as controversial (see e.g. \cite{Elli}),
it should be noted that it comes from the most sensitive ($^{76}$Ge)
detector to date and no independent experiment can, at the moment,
falsify it.

As very well known in the literature (see e.g.
\cite{pastor}) massive neutrinos
can be extremely relevant for cosmology and leave key signatures
in several cosmological data sets. More
specifically, massive neutrinos suppress the growth of fluctuations on
scales below the horizon when they become non relativistic.
\noindent Current cosmological data, in the framework of
a cosmological constant, are able to indirectly constrain the
absolute neutrino mass to $\Sigma m_{\nu} < 0.75$ eV at $95 \%$ c.l.
\cite{wmap3cosm} and are in tension with the HM claim.
However, as first noticed by \cite{hannestad}, there is some
form of anticorrelation between the equation of state parameter
$w$ and $\Sigma m_{\nu}$.
The cosmological bound on neutrino masses can therefore be relaxed
by using a DE component with a more negative
value of $w$ than a cosmological constant. As we show here, the HM claim
is compatible with the cosmological data only if
the equation of state (parameterized as constant) is
$w < -1$ at $95 \%$.

While the HM claim must certainly verified by future experiments
there are several theoretical motivations that one should consider
before discarding the result as simply due to unaccounted
systematics. In the past years, indeed, the interesting idea of a
possible link between neutrinos and DE has been proposed \cite{IDE}.
The main motivation for this connection relies on the fact that the
energy scale of DE (${\cal  O}(10^{-3})$~eV) is of the order of the
neutrino mass scale. Interestingly enough, coupled neutrino-DE
models \cite{IDE} predict in general an effective equation of state
(averaged over redshift) $w_{eff} < -1$. Our result may therefore be
the consequence of a deeper connection between DE and neutrino
physics.

The method we adopt is based on the publicly available Markov Chain Monte Carlo
package \texttt{cosmomc} \cite{Lewis:2002ah} with a convergence
diagnostics done through the Gelman and Rubin statistic.
We sample the following eight-dimensional set of cosmological
parameters,
adopting flat priors on them: the physical baryon, Cold Dark Matter
and massive neutrinos densities,
$\omega_b=\Omega_bh^2$,
$\omega_c=\Omega_ch^2$ and $\Omega_{\nu}h^2$,
 the ratio of the sound horizon to the angular diameter
distance at decoupling, $\theta_s$, the scalar spectral index $n_s$,
the overall normalization of the spectrum $A$ at
$k=0.05$ Mpc$^{-1}$, the optical
depth to reionization, $\tau$, and, finally, the
DE equation of state parameter $w$.
Furthermore, we consider purely adiabatic initial conditions and we
impose flatness.

We include the three-year WMAP data \cite{wmap3cosm} (temperature
and polarization) with the routine for computing the likelihood
supplied by the WMAP team. Together with the WMAP data we also
consider the small-scale CMB measurements of CBI
\cite{2004ApJ...609..498R}, VSA \cite{2004MNRAS.353..732D}, ACBAR
\cite{2002AAS...20114004K} and BOOMERANG-2k2
\cite{2005astro.ph..7503M}.  In addition to the CMB data, we include
the constraints on the real-space power spectrum of galaxies from
the SLOAN galaxy redshift survey (SDSS) \cite{2004ApJ...606..702T}
and 2dF \cite{2005MNRAS.362..505C},
 and the Supernovae Legacy Survey data from \cite{2006A&A...447...31A}.
Finally, we include the Heidelberg-Moscow as in the recent analysis of
\cite{fogli2}.

Let us just remind that the $0\nu2\beta$ decay half-life
$T^{0\nu}_{1/2}$ is linked to the effective Majorana mass
$m_{\beta\beta}$ by the relation
$m^2_{\beta\beta}=m^2_e/C_{mm}T^{0\nu}_{1/2}$, in the assumption
that the $0\nu2\beta$ process proceeds {\em only\/} through light
Majorana neutrinos and where the nuclear matrix element $C_{mm}$
needs to be theoretically evaluated. Using the theoretical input for
$C_{mm}({}^{76}\mathrm{Ge})$ from Ref.~\cite{faessler}, the
$0\nu2\beta$ claim of \cite{Kl04} is transformed in the
$2\sigma$ range
\begin{equation}
\log_{10}(m_{\beta\beta}/\mathrm{eV})= -0.23\pm 0.14
\label{logmbb2}\ ,
\end{equation}
i.e., $0.43<m_{\beta\beta}<0.81$ (at $2\sigma$, in eV).

Considering all current oscillation data (see \cite{fogli2}) and
under the assumption of a $3$ flavor neutrino mixing the above
constraint yields:
\be\la{OmegaNeu}
0.0137< \Omega_{\nu}h^2 <0.026
\ee at
$95 \%$ c.l.
where we used the well known relation:
${\Omega_{\nu}h^2}=\Sigma m_{\nu}/93.2 eV.$

Our main results are plotted in Fig.1 where we show the constraints on
the $w-\Sigma m_{\nu}$ plane in two cases,
 with and without the HM prior on
neutrino masses. As we can see, without the
HM prior we are able to reproduce the results already
presented in the literature (see e.g. \cite{wmap3cosm}),
namely current cosmological data constrain  neutrino masses
to be $\Sigma m_{\nu} <0.75\, eV$.
However an interesting anti-correlation is present between the DE
parameter $w$ and the neutrino masses and larger
neutrino masses are in better agreement with the data
for more negative values of $w$.
It is therefore clear that when we add the HM prior
($\Sigma m_{\nu} \sim 1.8 \pm 0.6\,eV$ at $95 \%$ c.l., again see Fig.1)
the contours are shifted towards higher
values of neutrino masses and towards lower values
of $w$. A combined analysis of cosmological
data with the HM priors gives
$-1.67< w <-1.05$
and $0.66<\Sigma m_{\nu}<1.11$ (in eV)
at $95 \%$ c.l. excluding the case of the cosmological constant
at more than $2 \sigma$ with
$\Sigma m_{\nu} = 0.85\,eV$, $w =-1.31  $ and $\Omega_m=0.35$
as best fit values.
Without the HM prior the data gives $-1.28 < w < -0.92$
and $\Sigma m_{\nu} < 0.73\,eV$
again at $95 \%$ c.l. with $w=-1.02$, $\Sigma m_{\nu} = 0.05\,eV$
and $\Omega_m=0.29$  as best fit
 values.

The inclusion of the HM prior affects also
other parameters. We found, at $95 \%$ c.l.:
  $0.916 < n_s < 0.979$
($0.926 < n_s < 0.989$ withouth HM), $0.0209 < \Omega_bh^2 < 0.0235$
($0.0211 < \Omega_bh^2 < 0.0238$ without HM),
$0.302 < \Omega_m < 0.444$ ($0.262 < \Omega_m < 0.360$ without HM).
It is interesting to notice that the inclusion of massive
neutrinos seems to further rule out the scale-invariant $n_{s}=1$
model.
\begin{figure}[tbp]
    \includegraphics[width=8.0cm]{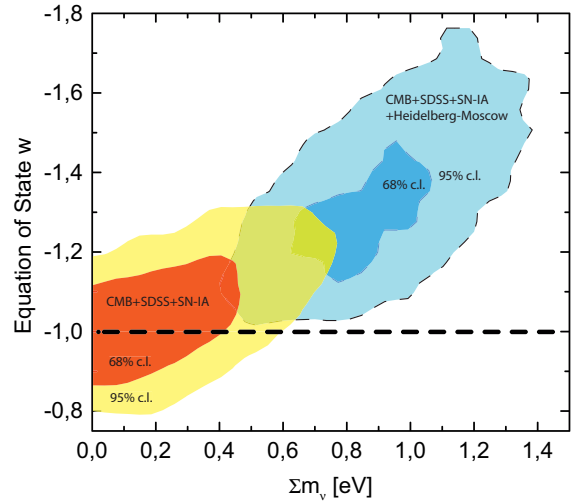}
  \caption{Constraints on the $w-\Sigma$ plane in two cases
with and without the Heidelberg-Moscow prior on
neutrino masses.}
 \label{fig1}
\end{figure}

As we have seen the combination  of a neutrino mass $\Sigma m_{\nu}
\simeq 0.85\, eV$ and the  cosmological observations indicate that
the equation of state of DE $w$ is less than than $-1$,
excluding a cosmological constant.
Scalar fields with positive kinetic energy have $w>-1$
while phantom fields \ci{phantom} can have $w<-1$
but they have a negative kinetic energy and
many fundamental theoretical problems.
Here, we prefer to follow the approach of interacting
 DE \cite{IDE}. Interacting DE are models where the dark
 energy interacts with other particles, as for example dark
 matter or neutrinos.
 The net effect of this interaction is to change  the apparent
 equation of state of DE
 \ci{IDE}. An observer that supposes
 that the DE has no interaction sees a different
 evolution of DE as an observer that takes into account
for the interaction of DE. This effect allows to have an  apparent
equation of state $w<-1$ for the ``non-interaction" DE observer even though
the true equation of state of the DE is larger than -1.

To sees this more clearly let us
define the energy density and pressure of the scalar field as
$
\rp=\fr{1}{2}\dot\phi^2+V(\phi)$,
$p_\phi=\fr{1}{2}\dot\phi^2-V(\phi)
$
and the equation of state parameter
$
\wp=p_\phi/\rp$,
where the potential $V(\phi)$ does not include the interaction with
dark matter. If there is no interaction between DE
and dark matter, $\wp$ gives the complete evolution of DE
and $w> -1$ .   We  now include an interaction term between
dark matter (or neutrinos) with $\phi$ via the function $f(\phi)$
which gives an interacting dark matter energy density \cite{Amendola:1999qq}
\be
\rim=\rimo\fr{f(\phi)}{f_o}\fr{1}{a^3}
\ee
where $f_o\equiv f(\phi_o)$ and $a_o=1$ at present time. In this
case dark matter no longer redshifts as $a^{-3}$ since the evolution
of $f(\phi)$ will also contribute to the redshift. The evolution
of $\rim$ and $\phi$ are given  by
\bea
\dot\rim+3H(\rim+p_{DM})&=& \fr{\rimo}{a^3}\fr{f'}{f_o} \; \dot\phi \\
\ddot\phi+3H\dot\phi +V'&=&-\fr{\rimo}{a^3}\fr{f'}{f_o}
\eea
 where the prime denotes derivative w.r.t. $\phi$, i.e. $V'\equiv dV/d\phi,\,f'\equiv df/d\phi$.
The total dark matter does not need to coincide with $\rim$. This
would be the case if we want to interpret $\rim$ as the energy density of
neutrinos since we know that they cannot give the total amount of dark
matter. However, since neutrinos are massive  they certainly contribute
to dark matter.

It was pointed out in \ci{IDE} that the apparent equation of state, i.e. the
equation of state of DE if we had assume that there was no interaction,
is
\be\la{wap}
w_{ap}=\fr{\wp}{1-x},\hspace{.5cm}
x=-\fr{\rimo}{\rp\,a^3}\le(\fr{f(\phi)}{f_o}-1\ri).
\ee
In this case the noninteracting DE and dark matter observer
sees a standard  evolution, $\dot\rho_{m}=-3H\rho_{m}$ and
 $\dot\rho_{DE}=- 3 H \rho_{DE}(1+w_{ap})$.
We see from eq.(\ref{wap})
that for $f<fo$ we have $x>0$ and $w_{ap} < \wp$, which allows to have
a $w_{ap}$ less than -1.

The effect of an apparent $w_{ap}$ less than -1 has a stronger effect
on small redshifts when the DE dominates. This effect is
measured by the SNIa and the actual values for the redshifts of these
supernovae are mostly in the range  $0 < z < 1.2$. So, let us
expand the function $f(\phi(a))$ as a function of the scale factor
around $a_o=1$,
\be
f(\phi)= f_o+\le(\fr{d f}{d\phi}\fr{d \phi}{da}\ri)|_{a_o} (a-1) + \ldots
\la{f}\ee
For generality  and presentation purposes  we assume that the scalar field
is already tracking, i.e. we take $\wp$ constant, and then the energy
density is given by $\rp=\rpo a^{-3(1+\wp)}=(2/(1-\wp)) V(\phi)$, where
we have used that the kinetic energy can be expressed as $E_k=(1+\wp)/(1-\wp) V$.
Taking the derivative
of $\rp$ w.r.t. $a$ we can relate  $d\phi/da$ to the potential $V$,
 $d\phi/da=-3(1+\wp)V/(a\, V')$. Using this expression of $d\phi/da$,
$\rp$ as a function of $a$  and  eq.(\ref{f}),  we
 can write $x$ in eq.(\ref{wap}), in terms of redshift z=1/a-1,  as
\be
 x=3\,A\;\left(\fr{1+\wp}{\Ompo}\right)\;\fr{z}{ (1+z)^{3\wp}}
\la{x2}
\ee
 where we have defined $ A\equiv - \Oimo \lmim/\lmp$ with $\lmim\equiv f'_o/f_o,\;
 \lmp\equiv V'_o/V_o$. A positive $x$ requires $A>0$.
 Notice that the evolution of $w_{ap}$ and $x$ depends on
 $z$ only via the term $z/(1+z)^{3\wp}$ in eq.(\ref{x2}) and once $\wp$ and
 $\Ompo$ are fixed the value of $w_{ap}$ is determined by
 the present day values of $\Oimo,f_o,V_o,f'_0,V'_o$ only through $A$.

Since $w_{ap}$ is a function of $z$ it is better to use the weighted
average equation of state to compare the models with the observational data.
The average equation of state is defined by
\be \la{weff}
w_{eff}=\fr{\int   w_{ap}\Omde \;dz}{\int \Omde  \; dz}
\ee
where the integral runs from $z=0$ to $z=1.2$. The effective
$w_{eff}$ is then a function of  $\Ompo$,   $\wp$ and $A$.

\begin{figure}[tbp]
   \includegraphics[width=8.0cm]{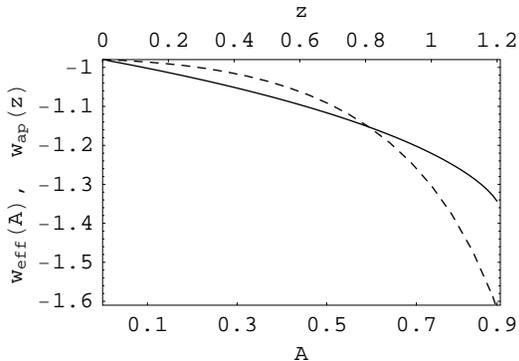}
  \caption{We show $w_{ap}$ as a function of z (dashed line)
  and $w_{eff}$ as a function of $ A\equiv-\Oimo\lmim/\lmp$ (solid line)
  with integration limits $0\leq z \leq 1.2$ for $\Ompo=0.65,\,\wp =-0.98$. }
 \label{grwap}
\end{figure}

We show in Fig.(\ref{grwap}) the evolution of $w_{ap}$ as
 a function of $z$ (dashed line) for $\Ompo=0.65, A=0.35$ (i.e.
 $\Oimo=0.35$ if $\lmim/\lmp=-1$), $\wp=-0.98$. We see that
 $w_{ap}$ decreases with increasing $z$ and becomes less than -1 at $z=0.3$
 and it is $w_{ap}=-1.6$ at $z=1.2$. We also show
in Fig.(\ref{grwap}) the behavior of $w_{eff}$ as a function of $A$ (solid line)
with the same parameters $\Ompo=0.65,\, \wp=-0.98$. With increasing values of $A$,
 $w_{eff}$ becomes more negative and for $A=0.1$
we find
$w_{eff}=-1$ and at $A=0.84$ we have $w_{eff}=-1.3$ as required by
the cosmological plus HM data. Finally, if we  assume that the interacting matter
is only due
to neutrinos with the total amount of neutrinos today given by the central values
of the CMB plus HM analysis
$\Sigma m_{\nu} =0.85\,eV$ then $\Oimo h^2= \Omega_\nu h^2=0.009$ and
$\lmim/\lmp=-40$ for $w_{eff}=-1.3$.

We have seen in a model independent study that using interacting DE
it is possible to obtain $w_{eff}$ less than -1,  consistent with
the values given by the cosmological data plus HM. Future high -z
baryon acoustic oscillation surveys, in tandem with high-z
supernovae surveys should provide a powerful mechanism to look for
such deviations from $w = - 1$.

To summarize and conclude,
 we have considered in this {\it letter} the cosmological implications of
the controversial Heidelberg-Moscow result.
We have found that a scenario based on a cosmological
constant is unable to provide a good fit to current data when
a massive neutrino component as large as suggested by HM is included
in the analysis. A better fit to the data is obtained when
the DE component is described with an equation of state
$w \sim -1.3$, with $w < -1$ at more than $95 \%$ c.l..
As far as we know, this is the only dataset able to exclude
a cosmological constant at such high significance.

There exists, therefore, a significant tension between the indirect,
observational measurements leading to the LCDM scenario and the direct
HM observations. Rather than implying one should rule out evidence
from the direct measurements purely on the basis of disparity with
the indirect observations, this tension suggests we should keep
our minds open to alternative dark energy scenarios beyond a
cosmological constant. This, together with the fact that the energy
scale of DE (${\cal  O}(10^{-3})$~eV) is of the order of the
neutrino mass scale, may suggest for a link between neutrino
physics and DE that must certainly be further investigated.
Systematics can be present in the HM data and a more conservative
treatment (see \cite{Kl04}) would lead
to a better agreement with a cosmological constant.
However, phantom models with $w < -1$ would still provide
a better fit to the data.
On the other hand, using a more conservative
approach towards cosmology, by, for example, combining HM only with
the CMB dataset, would provide even larger values of $\Sigma m_{\nu}$
and more negative values for $w$.
Recent combined analysis with Lyman-$\alpha$ forest data
(\cite{seljakanze},\cite{fogli2})
imply tight constraints on neutrino masses ($\Sigma m_{\nu} <0.2eV$),
seemingly at discord with the HM result, and also in some tension with CMB data alone.
 Future larger scale Lyman-$\alpha$
surveys and refinements in the analysis, addressing systematic
uncertainties and sensitivity to modeling assumptions,
will allow a better assessment of how these tensions will be resolved.

If the HM result is correct, a signal in the range
 $m_{\nu}\sim\mathrm{few}\times 10^{-1}$ eV is clearly expected, and could be
 found in the next-generation Karksruhe Tritium Neutrino Experiment
 (KATRIN) \cite{katrin}, which should take data in the next decade,
 with an estimated sensitivity down to $\sim 0.2$ eV.
Cosmological data will also reach a $2\sigma$ accuracy
of about $\sim 0.1$ eV on $\Sigma m_{\nu}$ in the
future \cite{pastor} mostly thanks to
Planck satellite CMB experiment expected to launch in 2008,
and future precision high-z supernovae and baryon oscillation
dark energy surveys.

A determination of the absolute neutrino mass scale will therefore
not only bring relevant information for neutrino physics but may be
extremely important in the determination of the dark energy
properties and in shedding light on a possible neutrino-dark energy
connection. Future direct particle detection and indirect astronomical
 experiments will scrutinize this interesting hypothesis.

{\it Acknowledgments}
It is a pleasure to thank Elio Lisi and Antonio Palazzo.
This work was made possible by the partial support of  ALFA-EC funds
in the framework of the HELEN Project,
CONACYT 45178-F and UNAM, DGAPA IN114903-3 projects.
RB is supported by NSF grant AST-0607018.

\end{document}